\documentclass[10pt]{article}
\usepackage{latexsym}
\newcommand{\be}{\begin{equation}}
\newcommand{\ee}{\end{equation}}

\catcode `\@=11
\catcode `\@=12
\begin{document}
\begin{center}
\large{\bf {CAUSAL BULK VISCOUS LRS BIANCH I MODELS WITH VARIABLE 
GRAVITATIONAL AND COSMOLOGICAL ``CONSTANT''}}\\ 
\vspace{10mm}
\normalsize{Anirudh Pradhan$^1$, Purnima Pandey$^2$, G. P. Singh$^3$ and R. V. Deshpandey $^4$} \\
\vspace{5mm}
\normalsize{$^1$ Department of Mathematics, Hindu Post-graduate College,
 Zamania-232 331, Ghazipur, India.} \\
\normalsize{E-mail : acpradhan@yahoo.com, pradhan@iucaa.ernet.in}\\
\vspace{5mm}
\normalsize{$^2$ Department of Mathematics, Hindu Post-graduate College,
 Zamania-232 331, Ghazipur, India.} \\
\normalsize{E-mail : purnima\_pandey2001@yahoo.com}\\
\vspace{5mm}
\normalsize{$^3$ Department of Mathematics, Visvesvaraya N. I. T.,
Nagpur-440 001, India} \\
\normalsize{E-mail : gpsingh@vnitnagpur.ac.in}\\
\vspace{5mm}
\normalsize{$^4$ Department of Mathematics, Priyadarshini College of 
Engineering and Architecture, Nagpur-440 016, India.} \\
\normalsize{E-mail : rvdpcea@yahoo.co.in}\\
\end{center}
\vspace{10mm}
\begin{abstract} 
In this paper we have investigated an LRS Bianchi I anisotropic cosmological model 
of the universe by taking time varying $G$ and $\Lambda$ in the presence of bulk 
viscous fluid source described by full causal non-equilibrium thermodynamics. We
obtain a cosmological constant as a decreasing function of time and for $m, n > 0$,
the value of cosmological ``constant'' for this model is found to be small and positive
which is supported by the results from recent supernovae observations. 
\end{abstract}
\smallskip
\newpage
\section{Introduction}
The conventional theory of evolution of the universe includes a number of dissipative 
processes. Dissipative thermodynamics processes in cosmology originating from a bulk 
viscosity are believed to play an important role in the dynamics and evolution of
the universe. Misner\cite{ref1} suggested that large-scale isotropy of the universe
observed at the present epoch is due to action of neutrino viscosity which was not
negligible when the universe was less than a second old. A number of processes 
responsible for producing bulk viscosity in the very early universe are such as the 
interaction between radiation and matter\cite{ref2}, gravitational string production 
\cite{ref3,ref4}, viscosity due to quark and gluon plasma, dark matter or particle
creation\cite{ref5,ref6}. It is important that each dissipative process is subject to 
as detailed analysis as possible. However, it is also important to develop a robust 
model of dissipative cosmological processes in general, so that one can analyze the 
overall dynamics of dissipative without getting lost in the details of particular
complex processes. In requirements of such a model Maartens\cite{ref7} pointed out
that the model should (i) be causal and stable, and (ii) provide a constant relativistic 
thermodynamics in the 'conventional' post-inflationary regime.\\
\newline
\par  
In order to study these phenomena, the theory of dissipative was first developed by 
Eckart\cite{ref8} and subsequently an essential equivalent formulation was given by
Landau and Lifshitz\cite{ref9}. But Eckart theory has several drawbacks including
violation of causality and instabilities of equilibrium states. Readers interested
in the general theory of causal thermodynamics are urged to consult the excellent
survey report of Maartens\cite{ref7} and Zimdahl\cite{ref10} and references cited 
therein. A relativistic second-order theory was found by Israel\cite{ref11} and
developed by Israel and Stewart\cite{ref12}. The advantages of the causal theory 
are as follows\cite{ref13}: (1) For stable fluid configurations, the dissipative signals 
propagate causally. (2) Unlike Eckart-type's theory, there is no generic short-
wavelength secular instability in causal theory. (3) Even for rotating fluids, 
the perturbations have a well-posed initial value problem. Therefore, the best 
currently available theory for analyzing dissipative processes in the universe 
is the full (i.e. non-truncated) Israel-Stewart causal thermodynamics, which we 
consider in this work.\\ 
\newline
\par   
In recent years, models with relic cosmological constant $\Lambda$ have drawn 
considerable attention among researchers for various aspects such as the age
problem, classical tests, observational constraints on $\Lambda$, structure
formation and gravitational lenses have been discussed in the literature (see
Refs.\cite{ref14}$-$\cite{ref16}). Lindey\cite{ref17} has suggested that cosmological 
``constant'' may be considered as a function of temperature and related to the
spontaneous symmetry breaking process. Therefore, $\Lambda$ should be a function
of time in a homogeneous universe as temperature varies with time. Some of the 
recent discussions on the cosmological constant ``problem'' and on cosmology with 
a time-varying cosmological constant are given by Ratra and Peebles \cite{ref18},
Dolgov\cite{ref19}$-$\cite{ref21}, Sahni and Starobinsky\cite{ref22}, Peebles\cite{ref23},
Padmanabhan\cite{ref24} and Vishwakarma\cite{ref25}. Recent cosmological observations
suggest the existence of a positive cosmological constant $\Lambda$ with the
magnitude $\Lambda(G \hbar/c^{3}) \approx 10^{-123}$. There are several aspects
of the cosmological constant both from cosmological and field theoretical 
perspectives. Presently, determination of $\Lambda$ has become one of the main 
issues of modern cosmology as it provides the gravity vacuum state and make possible
to understand the mechanism which led the early universe to the large scale structures
and to predict the fate of the whole universe. The cosmological ``constant'' can be 
measured by observing quasars whose light gets distorted by gravity of galaxies
that lies between the quasars and Earth. Krauss and Turner\cite{ref26} have mentioned
that as $\Lambda$ term dominates the energy density of the universe, cosmologists
are correct in their attempt to evoke it once again for better understanding of
both the universe and fundamental physics.\\
\newline
\par
In the last few decades there have been numerous modifications of general relativity
in which gravitational ``constant'' $(G)$ varies with time\cite{ref27}. Considering
the principle of absolute quark confinement, Der Sarkissian\cite{ref28} has suggested 
that gravitational and cosmological ``constant'' may be considered as functions of
time parameter in Einstein's theory of relativity. A number of authors\cite{ref29}$-$\cite{ref40}
have considered time-varying $G$ and $\Lambda$ within the framework of general relativity.\\
\newline
\par
Motivated by the fact that bulk viscosity, gravitational and cosmological ``constants''
, are more relevant during early stages of the universe, in this paper, we have considered
the evolution of a LRS Bianchi I model with bulk viscous fluid in full causal non-equilibrium
thermodynamics, in presence of time-varying gravitational and cosmological ``constants''.\\ 
\section{THE Basic Equations}
A locally rotationally symmetric (LRS) Bianchi I space-time may be represented
by the line element
\begin{equation}
\label{eq1}
ds^{2} = dt^{2} - A^{2} dx^{2} - B^{2} (dy^{2} + dz^{2}),
\end{equation}
where metric potentials A and B are depending on cosmic time  $t$ only.\\
The Einstein's field equations with variable $G$ and $\Lambda$ are given by
\begin{equation}
\label{eq2}
R_{ij} - \frac{1}{2} R g_{ij} + \Lambda(t) g_{ij} = - 8\pi G(t) T_{ij}, 
\end{equation}
where $R_{ij}$ is the Ricci tensor; $R$ = $g^{ij} R_{ij}$ is the
Ricci scalar; and $T_{ij}$ is the energy-momentum tensor of cosmic fluid in the 
presence of bulk viscosity  given by 
\begin{equation}
\label{eq3}
T_{ij} = (\rho + p + \Pi)u_{i}u_{j} - p g_{ij}.  
\end{equation}
Here $\rho$, $p$ and $\Pi$ are the energy density, equilibrium pressure and 
bulk viscous pressure respectively and $u^{i}$ is the flow vector satisfying the 
relations $u^{i}u_{i} = 1$.\\
The Einstein's field equations (\ref{eq2}) for the line element (\ref{eq1})
lead to the following set of equations
\begin{equation}
\label{eq4}
\frac{2\dot{A}\dot{B}}{AB} + \frac{\dot{B}^{2}}{B^{2}} = 8\pi G(t)\rho + \Lambda(t),
\end{equation}
\begin{equation}
\label{eq5}
\frac{\ddot{B}}{B} + \frac{\dot{A}\dot{B}}{AB} + \frac{\ddot{A}}{A} = 
- 8\pi G(t)[p + \Pi] + \Lambda(t),
\end{equation}
\begin{equation}
\label{eq6}
\frac{2\ddot{B}}{B} + \frac{\ddot{B}^{2}}{B^{2}} = - 8\pi G(t)[p + \Pi] + \Lambda(t).
\end{equation}
A combination of equations (\ref{eq4}) - (\ref{eq6}) yield the continuity equation
\begin{equation}
\label{eq7}
8\pi \dot{G}\rho + 8\pi G\left[\dot{\rho} + (\rho + p + \Pi)\left(\frac{\dot{A}}{A} 
+ \frac{2\dot{B}}{B}\right)\right] + \dot{\Lambda} = 0.
\end{equation}
The usual energy-momentum conservation equation $T^{ij}_{;j} = 0$ suggests
\begin{equation}
\label{eq8}
\dot{\rho} + (\rho + p + \Pi)\left[\frac{\dot{A}}{A} + \frac{2\dot{B}}{B}\right] = 0.
\end{equation}
From equations (\ref{eq7}) and (\ref{eq8}), we get
\begin{equation}
\label{eq9}
\dot{\Lambda} = -8\pi \dot{G} \rho.
\end{equation}
For the full causal non-equilibrium thermodynamics, the causal evolution
equation for bulk viscosity is given by\cite{ref7}
\begin{equation}
\label{eq10}
\tau \dot{\Pi} + \Pi = - \xi\left(\frac{\dot{A}}{A} + \frac{2\dot{B}}{B}\right)
- \frac{\epsilon}{2} ~ \tau ~ \Pi \left(\frac{\dot{A}}{A} + \frac{2\dot{B}}{B} 
+ \frac{\dot{\tau}}{\tau} - \frac{\dot{\xi}}{\xi} - \frac{\dot{T}}{T}\right),
\end{equation}
where $T \geq 0$ is the temperature, $\xi$ the bulk viscous coefficient and
$\tau \geq 0$ the relaxation coefficient for the transient bulk viscous effect
(relaxation time i.e. the time which system takes in going back to equilibrium 
once the divergence of the four velocity has been switched off). For $\tau = 0$,
equation (\ref{eq10}) gives evolution equation for the non-causal theory. For
$\epsilon = 0$, we get causal evolution equation for truncated theory, which 
implies a drastic condition on the temperature, while $\epsilon$ takes value 
unity in full causal theory.\\
\section{The Model}
Equations (\ref{eq5}) and (\ref{eq6}) yield
\begin{equation}
\label{eq11}
\frac{\ddot{B}}{B} + \frac{\dot{B}^{2}}{B^{2}} - \frac{\ddot{A}}{A} - 
\frac{\dot{A}\dot{B}}{AB} = 0.
\end{equation}
In order to find exact solutions of the field equations, we require more
physically plausible relations amongst the variables. Considering a power
law relation between $A$ and $B$ as $A \sim B^{n}$, eq.(\ref{eq11}) suggests
\begin{equation}
\label{eq12}
B = B_{0}\left(\frac{t}{t_{0}}\right)^{\frac{1}{n + 2}},
\end{equation}
\begin{equation}
\label{eq13}
A = A_{0}\left(\frac{t}{t_{0}}\right)^{\frac{n}{n + 2}},
\end{equation}
where $A_{0}$ and $B_{0}$ are the values of $A$ and $B$ at present time
$t = t_{0}$ and $n$ is a positive constant.\\
By use of (\ref{eq12}) and (\ref{eq13}), equations (\ref{eq4}) - (\ref{eq6}), 
reduce to
\begin{equation}
\label{eq14}
\frac{2n}{(n + 2)^{2}}\frac{1}{t^{2}} = 8 \pi G\rho + \Lambda,
\end{equation}
\begin{equation}
\label{eq15}
\frac{2n +1}{(n + 2)^{2}}\frac{1}{t^{2}} = 8 \pi G(p + \Pi) - \Lambda.
\end{equation}
Further following\cite{ref21}$-$\cite{ref25}, we assume $G = t^{m}$. Hence equations 
(\ref{eq9}) and (\ref{eq14}) suggest 
\begin{equation}
\label{eq16}
\dot{\Lambda} - \frac{m}{t}\Lambda = - \frac{2mn}{(n + 2)^{2}} \frac{1}{t^{3}},
\end{equation}
which is a linear equation and it has solution
\begin{equation}
\label{eq17}
\Lambda = \frac{2mn}{(n + 2)^{2}(m + 2)} \frac{1}{t^{2}}.
\end{equation}
From equations (\ref{eq14}) and (\ref{eq17}) one can easily obtain expression for
energy density in terms of cosmic time $t$ as
\begin{equation}
\label{eq18}
\rho = \frac{n}{2\pi (m + 2)(n + 2)^{2}} \frac{1}{t^{m + 2}}.
\end{equation}
Considering the usual barotropic equation of state relating the perfect fluid pressure
$p$ to the energy density as
\begin{equation}
\label{eq19}
p = (\gamma - 1)\rho,
\end{equation}
where $\gamma(1 \leq \gamma \leq 2)$ is a constant, equations (\ref{eq13}) and 
(\ref{eq14}) lead to
\begin{equation}
\label{eq20}
\Pi = \frac{n(m + 1)}{2\pi (m + 2)(n + 2)^{2}} \frac{1}{t^{2 + m}}.
\end{equation}
Now, we consider\cite{ref7,ref10}\cite{ref41}$-$\cite{ref44} following phenomenological widely 
accepted relations 
\begin{equation}
\label{eq21}
\xi - \xi_{0} \rho^{\alpha} ~ ~ and ~ ~ \tau = \frac{\xi}{\rho}
\end{equation}
for the bulk viscosity coefficient $\xi$ and mass density $\rho$ and also for
bulk viscosity coefficient and the relaxation time $\tau$, respectively, where 
$\xi_{0} \geq 0$ and $\alpha$ are constants. If $\alpha = 1$, eq.(\ref{eq21})
may correspond to a radiative fluid, whereas $\alpha = \frac{3}{2}$ may 
correspond to a string-dominated universe. However, more realistic 
models are based upon $\alpha$ in the region $0 \leq \alpha \leq \frac{1}{2}$. \\
Using (\ref{eq21}), equation (\ref{eq10}) on integration yields
\begin{equation}
\label{eq22}
T = T_{0}~ exp\left[\frac{2}{\epsilon} g(t)\right] ~ exp\left[\frac{2}{\epsilon} 
f(t)\right] \frac{\Pi^{\frac{2}{\epsilon}} R^{3} \tau}{\xi},
\end{equation}
where $f(t)$ and $g(t)$ are anti-derivatives of $\frac{1}{\tau}$ and 
$\frac{3\rho H}{\Pi}$ respectively.\\
With the help of equations (\ref{eq18}), (\ref{eq20}) and (\ref{eq21}), we 
obtain the expressions for $f(t)$ and $g(t)$ as
\begin{equation}
\label{eq23}
f(t) = \frac{\tau_{0}}{t^{(m + 2)(\alpha - 1) - 1}},
\end{equation}
\begin{equation}
\label{eq24}
g(t) = \frac{1}{m(m + 1)t^{m}},
\end{equation}
where
\begin{equation}
\label{eq25}
\tau_{o} = \frac{1}{1 - (m + 2)(\alpha - 1)}\left[\frac{n}{2\pi(m + 2)
(n + 2)^{2}}\right]^{\alpha - 1}.
\end{equation}
We observe from eq. (\ref{eq17}) that the cosmological constant in this model is
decreasing function of time and it approach a small value as time increases (
i.e., the present epoch). For $m, n > 0$ the value of cosmological ``constant''
for this model is found to be small and positive which is supported by the results 
from recent supernovae observations recently obtained by the High - z Supernova 
Team and Supernova Cosmological Project ( Garnavich {\it et al.};\cite{ref45} Perlmutter 
{\it et al.};\cite{ref46} Riess {\it et al.};\cite{ref47} Schmidt {\it et al.}
\cite{ref48}).\\
Using the observational values $\dot{G}/G = 10^{-11} yr^{-1}$ and $H_{p} = 
7.5 \times 10^{-11} yr^{-1}$, and the relation of the present age of the universe
with Hubble parameter ($H_{p}~t_{p}\sim \frac{2}{3}$), relation $G = t^{m}$
suggest
\[
m = \frac{2}{22.5}
\]
This clearly shows that gravitational parameter $G$ turns out to be an increasing 
function of time.\\
From eq (\ref{eq18}), it is observed that the energy density $\rho$ is decreasing
with evolution of the universe. For all $m > -2$, $n>0$, the energy conditions are 
satisfied.\\
It is also observed from eq (\ref{eq20}) that bulk viscous pressure $\Pi$ decreases
with time. When $t \rightarrow 0$, $\Pi \rightarrow \infty$. When $t \rightarrow 
\infty $, $\Pi \rightarrow 0$.\\ 
In this model, expressions for expansion and shear are
\begin{equation}
\label{eq26}
\theta = \frac{1}{t} ~ ~ and ~ ~ \sigma^{2} = \frac{2(n - 1)^{2}}{3 (n + 2)^{2} t^{2}}.
\end{equation}
\begin{equation}
\label{eq27}
\frac{\sigma^{2}}{\rho} = \frac{4\pi (n - 1)^{2} (m + 2)}{3n} t^{m}.
\end{equation}
Equation (\ref{eq27}) clearly shows the effect of time varying $G$ on the relative anisotropic.
For, $-2 < m < 0$, relative anisotropy is decreasing. Further, it can be observed that
$\sigma^{2} \propto \theta^{2}$, which indicate that the model does not approach isotropy for 
large value of $t$.\\  
\section*{Acknowledgements} 
A. Pradhan and G. P. Singh thank to the Inter-University Centre for Astronomy and Astrophysics, 
India for providing facility under Associateship Programmes where this work was carried out. \\
\newline
\newline

\end{document}